\documentclass[prl,nofootinbib,twocolumn,superscriptaddress]{revtex4}

\usepackage{amsmath,amssymb,graphicx} %use drftcite in draftmode otherwise
\usepackage{epsf}

\def \be  {\begin{equation}}
\def \ee  {\end{equation}}
\def \ba  {\begin{eqnarray}}
\def \ea  {\end{eqnarray}}
\def \baa {\begin{eqnarray*}}
\def \eaa {\end{eqnarray*}}
\def \bb  {\begin{thebibliography}}
\def \eb  {\end{thebibliography}}

\def \lab #1 {\label{#1}}

\newcommand \nn {\nonumber}
\newcommand \ts {\theta_s}

\def\pt{{p_{T}}}

\def \fracs #1#2 {\mbox{\small $\frac{#1}{#2}$}}

\def \bin #1#2 {{\left({#1}\atop{#2}\right)}}
\def \as {\relax\ifmmode\alpha_s\else{$\alpha_s${ }}\fi}

\def \al #1 {\frac {\as({#1})}{\pi} }
\def \ds #1 {\ooalign{$\hfil/\hfil$\crcr$#1$}}

\newcommand \bea{\begin{eqnarray}}
\newcommand \eea{\end{eqnarray}}

\def\hepph  #1 {{\tt hep-ph/#1}}

\begin{document}
\begin{titlepage}

\begin{flushright}
YITP-SB-08-31
\end{flushright}

\vskip.5cm
\begin{center}
{\large \bf  Substructure of high-$p_T$ Jets at the LHC}
\vskip.2cm
\end{center}

\begin{center}
{ {Leandro G. Almeida}$^a$, {Seung J. Lee}$^a$, {Gilad Perez}$^a$, {George Sterman}$^a$,  {Ilmo Sung}$^a$, {Joseph Virzi}$^b$} \\

\vskip 8pt

$^{a}$ {\it \small C. N. Yang Institute for Theoretical Physics\\
Stony Brook University, Stony Brook, NY 11794-3840, USA}\\

\vspace*{0.3cm}

$^b$ {\it \small Lawrence Berkeley National Laboratory\\
Physics Division, 1 Cyclotron Road, Berkeley, CA 94720, USA}
\end{center}

\begin{center} {\bf Abstract}\\\end{center}

We study high-$\pt$ jets from QCD and from highly-boosted
massive particles such as tops, $W,Z$ and Higgs, 
and argue that 
infrared-safe observables can help reduce QCD backgrounds.
Jets from QCD are characterized by different patterns
of energy
flow compared to the products of highly-boosted 
heavy particle decays, and
we employ a variety of {\it jet shapes}, observables
restricted to energy flow within a jet, to explore this difference.
Results from Monte Carlo generators and arguments based on
perturbation theory support 
the discriminating power of the shapes we refer to as {\em planar flow}
and {\em angularities}.   We emphasize that for massive jets, these and other
observables can be analyzed perturbatively.
 
\vskip .2in

\vbox{\vskip .25 in}
\end{titlepage}

{\bf Introduction.} 
At the Large Hadron Collider (LHC), events
with highly-boosted massive particles, tops~\cite{topjets}, $W$, $Z$ and
Higgs, $h$~\cite{scalarjets} may be the key ingredient for the 
discovery of physics beyond the
standard model
 \cite{ExpRef, Brooijmans:2008, Butterworth:2002tt}.   
 In many decay channels, these particles would be identified as
high-$p_T$ jets, and any such signal of definite mass must be disentangled from
a large background of light-parton (``QCD") jets with a continuous distribution.
This background far exceeds such signals, and relying solely on jet mass 
as a way to reject QCD background from signal would
probably not suffice in most cases \cite{us}, even using a
narrow window for dijets in the search for massive particles such as tops,
produced in pairs.   

In this paper, we argue that for  massive, high-$p_T$ jets, 
infrared (IR) safe observables
may offer an additional  tool to distinguish 
heavy particle decays from QCD background, perhaps even
on an event-by-event basis. 
We will refer to inclusive observables dependent on energy flow within
individual jets as {\em jet shapes}.
A jet within a cone of radius 0.4, for example, 
may be identified from the energy recorded in  roughly fifty 0.1$\times$0.1
calorimeter towers.
It thus contains a great deal of information.
Perturbatively-calculable, infrared safe 
jet shapes combined, of course, with IR-safe jet finding algorithms \cite{SIS},
may enable us to 
access that information systematically, and to form a bridge
between event generator output and direct theory predictions.  

Essentially, any observable that is a smooth functional of the distribution of energy flow
among the cells defines an IR-safe jet shape \cite{Sterman:1979uw}.
The jet mass is one example, but a single jet may be analyzed according to a variety of
shapes.
In particular, the jet mass distribution has large corrections
when the ratio of the jet mass to jet energy 
is small \cite{Kidonakis:1998bk}, but can
be computed at fixed order when the logarithm of that ratio
has an absolute value of order unity.
Once the jet mass is fixed at a high scale, 
a large class of other jet shapes become perturbatively
calculable with nominally small corrections. Indeed, a jet whose
mass exceeds ${\cal O}(100\,\rm GeV)$ becomes, from the point of view
of perturbation theory, much like the final state in
leptonic annihilation at a similar scale. At such
energies, event shapes, 
which in the terminology of this paper are jet shapes extended over all particles,
have been extensively 
studied in perturbation theory
\cite{Dasgupta:2003iq}.
In this study we explore the possibility that perturbative predictions for jet
shapes differ between those jets that originate from the decay of heavy particles, and those
which result from the showering of light quarks and gluons.
Very interesting related studies  have recently appeared 
in \cite{Brooijmans:2008, Kaplan:2008ie,Thaler:2008ju}.    

{\bf Jet Shapes and Jet Substructure.} 
We would like to identify 
jet shapes for which perturbative
predictions differ significantly between QCD and other high-$\pt$ jets,
focusing on relatively narrow windows in jet mass.  
In our companion paper~\cite{us} we have 
discussed how to calculate the jet mass distribution for the QCD background.
We now extend this argument to the computation of other 
jet shape observables.   

We emphasize that, because the observables under
consideration are IR-safe, we may calculate
them  as power series in $\alpha_s$, 
starting at first order for the QCD background, and zeroth order for an
electroweak decay signal.

Our approximation for the jet cross sections is based on
factorization for the relatively-collinear
partons that form a jet from the remainder of process \cite{Kidonakis:1998bk}.
For a jet of cone size $R$, contributions that do not vanish
as a power of $R$ are generated by a function that
depends only on the flavor of the parent parton, its
transverse momentum, and the factorization scale.
Denoting an jet shape by $e$, we then have,
\begin{equation}
\frac{d\sigma}{dm_J\,de}
=
\sum_c \int_{\pt^{min}}^\infty d\pt \frac{d\hat\sigma_c(\pt)}{d\pt}\ \frac{dJ_c(e,m_J,\pt,R)}{de}\, ,
\label{diffJ}
\end{equation}
where $d\hat\sigma/d\pt$ includes the hard scattering and the
parton distributions of the incoming hadrons, and where
the jet function for partons $c$ 
in the final state 
is defined formally as in Refs.\ \cite{us,Berger:2003iw}.   

In Ref.\ \cite{us},  we have found that the distribution
of QCD jet masses in the range of hundreds of GeV
is fairly well described by the jet function in Eq.\ (\ref{diffJ}) at order $\alpha_s$,
based on two-body final states.
It thus seems natural to anticipate that
for QCD jets, energy flow inside the cone would produce a {\emph{linear}} 
deposition in the detector \cite{Butterworth:2002tt,Thaler:2008ju, Brooijmans:2008, Kaplan:2008ie,Butterworth:2008iy}.
While this is certainly the case for an event consisting of two sub-jets, it is 
a simpler condition, and more easily quantified.
Indeed, such a linear flow 
should also characterize jets from the two-body decay of a highly-boosted, massive particle. 
We will see below that relatively simple jet shapes can help
distinguish QCD jets from many top-decay jets that involve three-body decay.
We will also see that 
jet shapes can help separate samples that contain both
QCD jets and jets from two-body decays, such as those of the $W$, $Z$ or $h$.
We emphasize that a single event may be analyzed by a variety of 
jet shapes, so that the resolution associated with each one need not be 
dramatic, so long as they are effectively independent.

{\bf Top decay and planar flow}.
The linear flow of QCD jets at leading order  
should be compared with a $\ge3$-body decay where the energy deposition 
tends to be {\emph{planar}},
covering a two-dimensional region of the detector.
An IR-safe jet shape, which we denote as {\emph{planar flow}},
a two-dimensional version of the ``$D$ parameter"
\cite{Parisi:1978eg, Donoghue:1979vi, Ellis:1980wv, Banfi:2001pb},  distinguishes planar from linear configurations.
 The utility of a closely-related observable was emphasized
 in Ref.\ \cite{Thaler:2008ju}.

Planar flow is defined as follows.
We first construct for a given jet a matrix  
$I_w$  as
\be
I^{kl}_{w}= {1\over m_J} \sum_i w_i \frac{p_{i,k}}{w_i}\,\frac{p_{i,l}}{w_i}\, ,
\ee
where $m_J$ is the jet mass, $w_i$ is the energy of particle $i$ in the jet,
and $p_{i,k}$ is the $k^{th}$ component of its transverse momentum relative to the 
axis of the jet's momentum.
Given $I_{w}$, we define $Pf$ for that jet as
\be
Pf={4\,{\rm det}(I_w)\over{\rm tr}(I_w)^2}={4 \lambda_1 \lambda_2\over(\lambda_1 + \lambda_2)^2}\, ,
\ee
where $\lambda_{1,2}$ are the eigenvalues of $I_w$.
$Pf$ vanishes for linear shapes and approaches unity for isotropic depositions of energy.

Jets with pure two-body kinematics have a differential jet function fixed at zero planar flow,
\be
\frac{1}{J}\, \left({dJ\over d Pf}\right)_{\rm 2 body}=\delta(Pf) \, .
\label{LOPf}
\ee
This would apply to leading order for events with highly boosted weak gauge boson, Higgs
and QCD jets.
On the other hand, events that are characterized by $\ge 3$-body kinematics have a 
smooth distribution.

Realistic QCD jets have,  of course,  nonzero $Pf$.
Because $Pf$ is an IR safe observable, however, its average value
can depend only on the hard
momentum scales of the jet, that is, $m_J$ and $p_T$.
This suggests an average $Pf$ of order $\alpha_s(m_J)\sim 0.1$
for high jet masses, times at most logarithms of that are order
unity for these heavy jets.
Correspondingly, higher orders corrections 
should, by analogy to two-jet event shapes  \cite{Dasgupta:2003iq},
replace the delta function of Eq.\ (\ref{LOPf})
with a differential distribution that peaks near the origin and then falls
off.   For jets resulting from three-body decay,
on the other hand, we anticipate that corrections in $\alpha_s$ 
shift the already-smooth distribution modestly, without affecting its overall shape.
Finally, for the vast majority of high-$\pt$ QCD jets, 
with masses $m_J\ll p_T$, planarity corrections associated with multi-gluon emission
may be expected to be large, and to shift $Pf$ to order unity.

\begin{figure}[t]
\begin{center}
 \includegraphics[width=.96085\hsize]{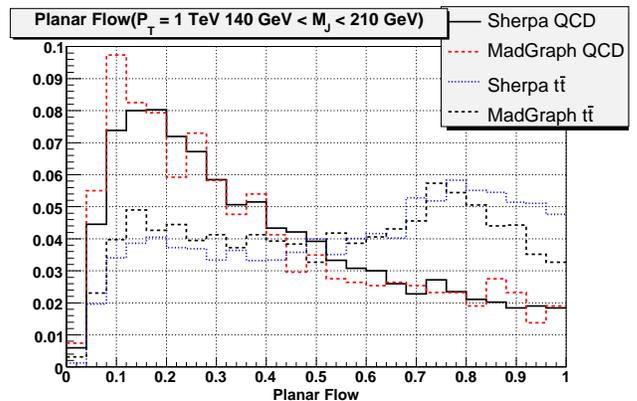}
 \ \\
 \end{center}
\caption{The planar flow distribution for QCD and top jets obtained from MadGraph and Sherpa. 
Distributions are normalized to unit area. 
 } \label{MCPl}
\end{figure}

These considerations are confirmed in Fig. \ref{MCPl},
where  we show the $Pf$ distribution
for QCD jet and hadronic $t\bar{t}$ events, for $R=0.4$, $\pt=1000\,$GeV and $m_J=140-210\,$GeV as obtained from MadGraph~\cite{MG} and Sherpa~\cite{SH} with jet reconstruction via 
(the IR-safe algorithm) 
SISCone~\cite{SIS}. 
We see that QCD jets peak around small values of $Pf$, while the top jet events are more dispersed.
A planar flow cut around 0.5 would clearly remove a considerably larger proportion of QCD jets than  top jets.
Correspondingly, we have confirmed by event generator studies that low-mass
QCD jets have much larger planar flows that those in Fig.\ \ref{MCPl}.

{\bf Two-body decay.}
While planar flow can help enrich samples with characteristically three- and higher-body
kinematics, other 
jet shapes can also provide additional information on events 
with relatively low $Pf$.  Here, we will still wish to distinguish the QCD background
from highly boosted electroweak gauge bosons or 
Higgs~\cite{Butterworth:2008iy} as well
as from top jets whose $Pf$ happens to  be relatively low.   
We begin with jets that are linear at lowest order, 
and identify a set of jet shapes
that have some power to distinguish between the two. Fixing $\pt,R$ and $m_J$ leaves only one free parameter to characterize the shape.

The QCD jet function for two-body kinematics is defined as a matrix element in
\cite{us} and is readily expressed as an integral over $\theta_s$,
the angle of the softer particle relative to the jet momentum axis.
For a quark jet, for example, the integrand is therefore the differential jet function, 
\bea
&&\frac{dJ^{QCD}}{d(c\ts)} = \frac{\alpha_s C_F \beta_z z^2}{4 \pi m_J^2 (1-\beta_z c\ts)(2(1-\beta_z c\ts)-z^2)}\times \nn \\
&&
\Bigg[ \frac{(2(1+\beta_z)(1-\beta_z c\ts)-z^2(1+c\ts))^2}{z^2(1+c\ts)(1-\beta_z c\ts)} +  3(1+\beta_z) +\nn \\
&&
 \frac{z^4(1+c \ts)^2}{(1-\beta_z c\ts)(2(1+\beta_z)(1-\beta_z c\ts)-z^2(1+c\ts))} \Bigg]\,, \label{djdtsQCD}
\eea
where $z\equiv m_J/\pt$, $\beta_z\equiv\sqrt{1-z^2}$ and $c\theta_s\equiv\cos\theta_s\,$.
The jet mass function is obtained by the integral $\int^{R}_{\theta_{m}} d\theta_s\, \left(dJ /d\theta_s\right)$,
where $\theta_m$ is the angle with the smallest possible value of the softest particle,
$\theta_m = \cos^{-1}\left({\sqrt{1-z^2}}\right)$,
at which both particles have the same energy and angle to the axis.

For signal events from a highly-boosted massive gauge bosons, 
we consider separately the 
cases when it is longitudinal and when it has helicity ($h=\pm 1$),
\bea &&\hspace*{-.4cm}
 {dJ^{\rm Long} \over d(c\theta_s)} = \frac{C}{(1-{\beta_z c\theta_s})^2}\, ,
 \label{djdts}\\
&&\hspace*{-.43cm}
{dJ^{h=\pm 1}\over d(c\theta_s)} = \frac{C}{(1-{\beta_z c\theta_s})^2} \left(1-\frac{(z s\ts)^2}{2(1-{\beta_z c\theta_s})^2} \right),
\nn 
\eea
where  $s\theta_s\equiv\sin\theta_s$ and $C$ is a proportionality coefficient, totally determined from the two-body decay kinematics.
We can interpret the appropriately normalized differential jet functions,
$P^x(\theta_s)=(d J^x /d \theta_s) /J^x$
as the probability to find the softer 
particle at an angle between $\theta_s$ and $\theta_s+\delta \theta_s$.
As the ratio $z$ decreases, the decay products become boosted and the
cone shrinks.   For QCD jets from light partons, however, this shrinkage is much less pronounced.
Plots of these jet functions show that the gauge boson distributions of Eq.\ (\ref{djdts})
fall off with $\theta_s$ faster than do QCD jets, Eq.\ (\ref{djdtsQCD}).
This observation suggests that
the signal (vector boson-jet) and background (QCD jets) have different shapes for fixed $\pt,R$ and jet mass.
This may be used to obtain an improved rejection power against background events.
We now consider a class of 
 jet shapes, {\emph{angularities}},
 originally introduced in
Ref.\ \cite{Berger:2003iw,Berger:2004xf} for two-jet events in $\rm e^+e^-$ annihilation.
A natural generalization of these 
jet shapes to single-cone jets of large mass $m_J$ is
\begin{equation}
\tilde\tau_a(R,\pt) = \frac{1}{m_J} \sum_{i \in jet} \omega_i\, \sin^a\left(\frac{\pi \theta_i}{2R}\right)\,
\left[ \,1 - \cos\left(\frac{\pi \theta_i}{2R}\right)\, \right]^{1-a}\, ,
\label{tauadef}
\end{equation}
with $m_J$ the jet mass.   The arguments of the trigonometric functions
vary from zero to $\pi/2$ as $\theta$ increases from zero to $R$, that
is, over the size of the cone.
These weights revert to the angularities as defined in
for leptonic annihilation in \cite{Berger:2003iw,Berger:2004xf} when
$R=\pi/2$, so that the cone is enlarged to a hemisphere and $m_J$ is
replaced by the center-of-mass energy in a two-jet event.
For massive jets, the angularities are clearly non-zero at lowest order,
in contrast to the lowest order planar flow, Eq.\ (\ref{LOPf}).   
Then, precisely because their IR safety, higher-order corrections
to the $\tau_a$ distributions should be moderate.

\begin{figure}[t]
\begin{center}
 \includegraphics[width=.85\hsize]{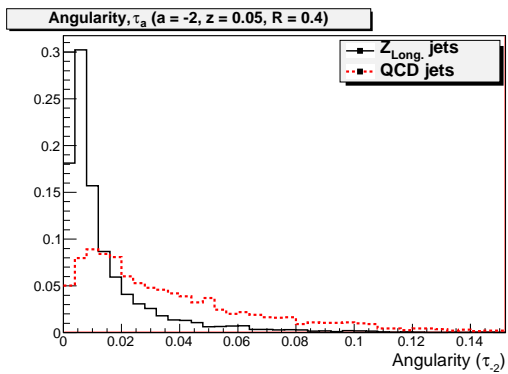}\\
 \end{center}
\caption{The angularity distribution for QCD (red-dashed
curve) and longitudinal $Z$ (black-solid curve) jets obtained from MadGraph. Both distributions are normalized to unit
area. 
 } \label{Angularity_Z_QCD}
\end{figure}

As the parameter $a$ varies, the angularities give more or less
weight to particles at the edge of the cone compared to those near the center.
\begin{figure}[th]
\begin{center}
 \includegraphics[width=.85\hsize]{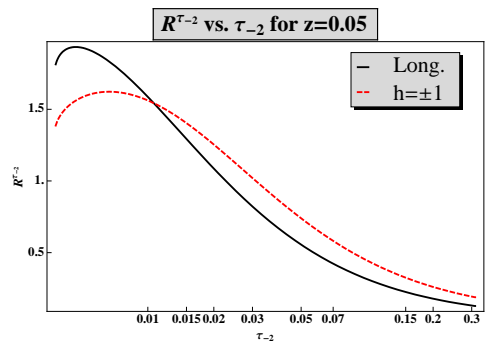}\\
 \end{center}
\caption{The ratio between the signal and background probabilities to have jet angularity  
$\tilde \tau_{-2}$, $R^{\tilde \tau_{-2}}$.
 } \label{Rtaua}
\end{figure}
From the differential jet distribution functions in Eqs.~(\ref{djdtsQCD}) and (\ref{djdts})
and the definition of $\tilde \tau_a$ we can obtain the expression for $P^x(\tilde\tau_a)$ 
[as before $x$= sig (signal) or QCD] the probability to find a jet with 
with an angularity value between $\tilde \tau_a$ and $\tilde \tau_a+\delta \tilde \tau_a$ at fixed $\pt,R,m_J$ and $a$. 
Our focus is not on the form of the individual distributions but rather on the ratio
of the signal to background
\begin{equation}
R(\tilde \tau_a)= \frac{P^{\rm sig}(\tilde\tau_a)}{ P^{\rm QCD}(\tilde\tau_a)} \, .
\label{RSB}\,
\end{equation}
In Fig.\ \ref{Rtaua} we show $R^{\tilde \tau_a}$ for  $a=-2$ and $z=0.05\,,$
for the different vector boson polarizations.    In Fig.\ \ref{Angularity_Z_QCD} we show the
corresponding angularity distributions
at the event generator level, comparing the output of MadGraph for longitudinal $Z$ boson
production to QCD jets in the same mass window.   
The pattern suggested by the lowest-order prediction
of Fig.\ \ref{Rtaua} is confirmed by the output of the event generator,
with signal and data curves crossing in Fig.\ \ref{Rtaua} near $\tau_{-2}=0.02$, 
where $R(\tilde\tau_{-2})\sim 1$.

{\bf Linear three-body decay.}
The leading-order differential top jet function can be obtained by considering
its three-body decay kinematics. 
The analytic expression is similar to Eq.\ (\ref{djdts}) for the two-body case, although a bit more elaborate.
In the following we simply point out a few features that  may help 
angularities to distinguish top jets from background, even
when they have relatively linear flow.

The lowest-order three-body
distribution is fully characterized by three angles. The first, $\theta_{b}$, is the angle between the $b$ quark and the 
jet axis.    The second, $\theta_{Wq}$, is the angle of the quark (from $W$-decay) relative to the $W$.
The third, $\phi$, is the angle of the same quark relative to the plane defined by the $W$ and the $b$.
For an on-shell $W$, the distributions peak around $\theta_b=\theta_m$ (as in two-body kinematics)
and $\theta_{Wq}= \theta_{m(W)}$ the minimal angle relative to the $W$ momenta in the $W$
rest frame.  Because it is massive, the  $W$'s decay products move in somewhat different directions, even in the
boosted frame, and their relative orientation induces the $\phi$-dependence.
Clearly, planar flow has maxima for odd multiples of $\phi=\pi/2$, and vanishes at lowest
order at multiples of $\pi$.    To tag top events
at zero planar flow, angularities can be 
of use.  
\begin{figure}[t]
\begin{center}
  \includegraphics[width=.85\hsize]{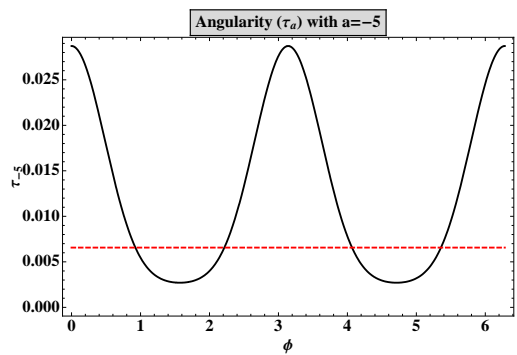}\\
 \ \\
 \end{center}
\caption{
Angularity, $\tilde \tau_{-5}$ as a function of the azimuthical angle of the $W(q\bar q)$ pair, $\phi_q$, for a typical top jet event, compared to the typical case two-body kinematics. 
 } \label{top_tau}
\end{figure}
In Fig.~\ref{top_tau} we plot  $\tilde \tau_{-5}$ as a function of the azimuthical angle of the $W(q\bar q)$ pair, $\phi$, for
a typical top jet event. We also show the corresponding value for the two-body case (clearly
$\phi$ independent). For illustration we choose the kinematical configuration that maximizes the corresponding differential jet distributions.
 We notice that this top angularity has
maxima with $\phi$ at zero and $\pi$ at values far above the most likely two-body configuration.
The reason is simply that angularities with 
large
negative values of $a$ tend to emphasize
flow at the edge of the cone.    Other values of $a$ weight individual jets differently
in general.   We consider this simple plot, along with the forgoing
examples from event generators, as strong evidence for the potential of jet shape analysis.

In summary, planar flow, angularities,
and jet shapes that are as yet to be invented, may afford a variety of tools with
which to distinguish the quantum mechanical histories of jets, whether resulting
from heavy particle decay, or strong interactions.

\subsection*{Acknowledgements}

The work of L.A., S.L., G.P., G.S.\, and I.S.\ was supported by the National Science Foundation, 
grants PHY-0354776, PHY-0354822, PHY-0653342 and PHY-06353354. 
The work of JV was supported by the Director, Office of Science, Office of High Energy Physics, of the U.S. Department of Energy under Contract No. DE-AC02-05CH11231.
We thank I. Hinchliffe and M. Shapiro for comments on the manuscript.

\end{document}